# Permanent Magnet Linear Generator Design for Surface Riding Wave Energy Converters

Farid Naghavi
*Dept. of Elec. & Comp. Eng*
*Texas A&M University*
College Station, USA
farid@tamu.edu

Shrikesh Sheshaprasad
*Dept. of Elec. & Comp. Eng*
*Texas A&M University*
College Station, USA
shrikesh@tamu.edu

Matthew Gardner
*Dept. of Elec. & Comp. Eng*
*University of Texas at Dallas*
Richardson, USA
matthew.gardner@utdallas.edu

Aghamarshana Meduri
*Dept. of Ocean. Eng*
*Texas A&M University*
College Station, USA
aghamarshana@tamu.edu

HeonYong Kang
*Dept. of Ocean. Eng*
*Texas A&M University*
College Station, USA
ga0prodg@tamu.edu

Hamid Toliyat
*Dept. of Elec. & Comp. Eng*
*Texas A&M University*
College Station, USA
toliyat@tamu.edu

*Abstract*— This paper describes the detailed analysis for the design of a linear generator developed for a Surface Riding Wave Energy Converter (SR-WEC), which was designed to improve energy capture over a wider range of sea states. The study starts with an analysis of the power take-off (PTO) control strategy to harness the maximum output power from given sea states. Passive, reactive, and discrete PTO control are explored. For the random wave excitation and limited sliding distance of the generator, the discrete strategy provides the highest average power output. The paper discusses the sizing requirement for the linear generator. Based on the force and power rating of the system and the application requirements, a slotless permanent magnet tubular generator is designed for the wave energy converter.

## I. Introduction

Ocean waves contain significant renewable energy, equivalent to 12.9 % of the U.S. Annual Energy Production (AEP) with a power density of 8 kW/m. This includes 60% of the West Coasts AEP and over 100% of Alaska's and Hawaii's AEP based on their 2012 electrical profiles [1]. The recently invented Surface Riding Wave Energy Converter (SR-WEC) [2] provides a new approach to competitively convert wave energy to renewable electricity in small or intermediate scales. As shown in Figure 1, the SR-WEC is uniquely designed to have the wave slopes excite tilting motion resonance. The relative invariance of the wave slopes throughout different sea states allows an inherently extended operating window in annual operation, and the rotational tilting motions make resonance control easier through relocating a mass [3]. When tilted, gravity causes the magnet assembly in the SR-WEC to slide along the center rod. A generator converts this kinetic energy to electrical energy. To ensure reliable long-term production with a simpler system [4], the SR-WEC uses a permanent magnet (PM) linear generator sealed inside the cylinder, which improves survivability beyond other existing wave energy converters with generation interfaces exposed to the ocean waves.

Research was sponsored by the Department of Energy under Grant No. DE-EE0008630

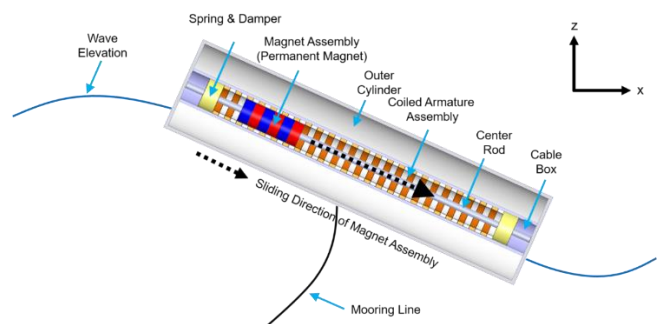

Figure 1: Surface Riding Wave Energy Converter [5].

The PM Tubular Linear (PMTL) generator has been recognized as a suitable candidate for wave energy converters [5]-[12]. A PM generator offers a higher force density and higher efficiency than other types of generators [6]. Various tubular PM generators have been proposed with radial, axial, and Halbach array magnet arrangements [7]-[12]. One challenge for tubular PM generators is the cogging force due to the stator teeth [11], [12], which can reduce the amount of power the SR-WEC is able to extract from the waves, especially in sea-states with small waves. To eliminate the cogging force, a slotless stator design has been adopted for the generator designed in this paper. In addition, to avoid cable stress and reliability issues due to movement, the stator windings are placed on the stationary part of the generator (Fig. 2). The length of the stator and, thus, the generator can be modularly increased according to the required length from power take-off (PTO) studies. However, only the overlapping region between the translator (magnets) and the stator winding produces generation force at any instant.

## II. Power Take-off Strategy and Generator Ratings

### A. Power Take-off Studies

The power take-off (PTO) strategy is responsible for ensuring that the WEC is utilized in the most effective manner by extracting as much electrical energy from the waves as possible. Various PTO damping strategies are discussed in [13]

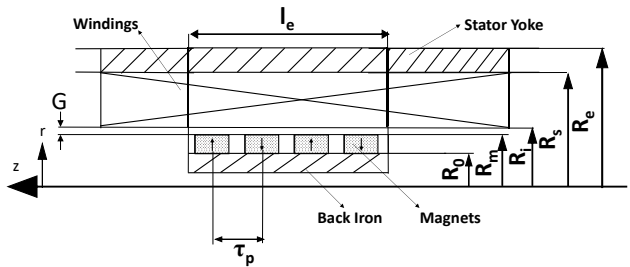

Figure 2: PMTL Generator Geometry based on [10].

$R_0$: Back iron outer radius
$R_m$: PM outer radius
$R_i$: Coil inner radius
$R_s$: Coil outer radius
$R_e$: Stator outer radius
$G$: Airgap
$l_e$: Translator length
$\tau_p$: Pole pitch

and [14]. These include passive, reactive and discrete PTO strategies. The intensity and duration of force applied during energy harvesting is set by the PTO strategy. Thus, these strategies play an important role in determining the generator specifications of the SR-WEC. These strategies include:

*1) Passive Damping*: Passive PTO damping replicates a simple viscous damping of the sliding motion. The force applied on the PM translator is directly proportional to its speed:

$$F_{PTO} = -C_{PTO}\dot{X}_{rel}, \quad (1)$$

where $F_{PTO}$ is the force applied by the generator, $C_{PTO}$ is the viscous damping coefficient, and $\dot{X}_{rel}$ is the speed of the sliding mass relative to the stator.

*2) Reactive Damping*: Reactive PTO damping replicates a viscous damper along with a stiffness coefficient (like a spring). Here, the force applied on the PM translator is directly proportional its speed and position:

$$F_{PTO} = -K_{PTO}X_{rel} - C_{PTO}\dot{X}_{rel} \quad (2)$$

where $K_{PTO}$ is the stiffness coefficient and $X_{rel}$ is the position of the sliding mass relative to the stator. Thus, the generator emulates both an electrical spring and an electrical viscous damper.

*3) Discrete Damping:* Here, the generator is always OFF (no force or power) or ON. Whenever the generator is ON, it generates as much instantaneous power as possible, subject to its force and power ratings. The generator is turned ON whenever the sliding mass approaches the end of the tube or when the tube changes its direction of tilt such that the mass is sliding uphill. The generator is then turned OFF when the sliding mass is brought to a stop.

*4) PTO Comparison:* A dataset of 11 random sea states with varying peak periods were used to compare the efficacy of The different PTO strategies. Wave spectral data is collected from National Data Buoy Center (NDBC) buoy #41002, located at a depth of 3920 m off the coast of Wilmington, North Carolina. The data consists of 7273 data points measured through 2018 [15]. The data is plotted as a resource characterization bin in Table I where each entry is the percentage of total data points that occurs in the given bin of significant wave height and energy period. The 11 peak periods, ranging from 4.06 s to 15.7 constitute 99.79% of all data points, representing a broad swathe of waves at the location. Using coupled time domain simulation of the SR-WEC, we obtained time series tilting motion data responding to respective random sea states and then solved the time domain sliding motions with the three PTO loads coupled. Fig. 3 shows the resulting tube slopes for a portion of the time. Table II shows the average power harvested using each control strategy, assuming the generator is 100% efficient but subject to peak force and power limits. The discrete PTO damping provides the most average power from the SR-WEC given the limited sliding distance for the PM translator inside the generator and the random waves.

TABLE I  RESOURCE CHARACTERISTIC BIN OF THE WAVE DATA

| | Occurrence % | Energy Period $T_e$(s) | | | | | | | | | | |
|---|---|---|---|---|---|---|---|---|---|---|---|---|
| | | 3.5 | 4.5 | 5.5 | 6.5 | 7.5 | 8.5 | 9.5 | 10.5 | 11.5 | 12.5 | 13.5 |
| | 0.25 | 0.00 | 0.00 | 0.07 | 0.18 | 0.23 | 0.00 | 0.00 | 0.00 | 0.00 | 0.00 | 0.00 |
| | 0.75 | 0.44 | 2.42 | 2.38 | 4.85 | 7.47 | 0.89 | 0.63 | 0.36 | 0.32 | 0.00 | 0.00 |
| | 1.25 | 0.61 | 5.57 | 8.11 | 3.88 | 8.84 | 1.28 | 1.10 | 0.30 | 0.36 | 0.00 | 0.01 |
| | 1.75 | 0.01 | 1.18 | 5.36 | 4.44 | 5.35 | 1.43 | 1.14 | 0.33 | 0.30 | 0.00 | 0.11 |
| | 2.25 | 0.00 | 0.15 | 2.13 | 3.69 | 2.43 | 0.39 | 0.67 | 0.15 | 0.23 | 0.10 | 0.01 |
| | 2.75 | 0.00 | 0.00 | 0.74 | 2.21 | 2.97 | 0.41 | 0.54 | 0.18 | 0.21 | 0.07 | 0.04 |
| | 3.25 | 0.00 | 0.00 | 0.10 | 1.27 | 2.43 | 0.44 | 0.23 | 0.11 | 0.06 | 0.12 | 0.06 |
| | 3.75 | 0.00 | 0.00 | 0.01 | 0.33 | 1.27 | 0.28 | 0.17 | 0.06 | 0.15 | 0.14 | 0.12 |
| Significant Wave Height $H_s$ (m) | 4.25 | 0.00 | 0.00 | 0.00 | 0.04 | 0.61 | 0.36 | 0.25 | 0.03 | 0.10 | 0.11 | 0.12 |
| | 4.75 | 0.00 | 0.00 | 0.00 | 0.00 | 0.33 | 0.25 | 0.26 | 0.00 | 0.03 | 0.14 | 0.15 |
| | 5.25 | 0.00 | 0.00 | 0.00 | 0.00 | 0.06 | 0.12 | 0.17 | 0.03 | 0.06 | 0.12 | 0.12 |
| | 5.75 | 0.00 | 0.00 | 0.00 | 0.00 | 0.10 | 0.11 | 0.12 | 0.04 | 0.08 | 0.07 | 0.03 |
| | 6.25 | 0.00 | 0.00 | 0.00 | 0.00 | 0.03 | 0.04 | 0.11 | 0.04 | 0.03 | 0.06 | 0.21 |
| | 6.75 | 0.00 | 0.00 | 0.00 | 0.00 | 0.00 | 0.03 | 0.03 | 0.06 | 0.07 | 0.07 | 0.14 |
| | 7.25 | 0.00 | 0.00 | 0.00 | 0.00 | 0.00 | 0.00 | 0.01 | 0.00 | 0.08 | 0.03 | 0.08 |
| | 7.75 | 0.00 | 0.00 | 0.00 | 0.00 | 0.00 | 0.00 | 0.00 | 0.00 | 0.01 | 0.03 | 0.00 |
| | 8.25 | 0.00 | 0.00 | 0.00 | 0.00 | 0.00 | 0.00 | 0.00 | 0.00 | 0.01 | 0.00 | 0.01 |
| | 8.75 | 0.00 | 0.00 | 0.00 | 0.00 | 0.00 | 0.00 | 0.00 | 0.00 | 0.00 | 0.01 | 0.00 |
| | 9.25 | 0.00 | 0.00 | 0.00 | 0.00 | 0.00 | 0.00 | 0.00 | 0.00 | 0.00 | 0.00 | 0.00 |
| | 9.75 | 0.00 | 0.00 | 0.00 | 0.00 | 0.00 | 0.00 | 0.00 | 0.00 | 0.00 | 0.00 | 0.00 |
| Peak Period $T_p$(s) | | 4.06 | 5.22 | 6.38 | 7.54 | 8.7 | 9.86 | 11.02 | 12.18 | 13.34 | 14.5 | 15.66 |

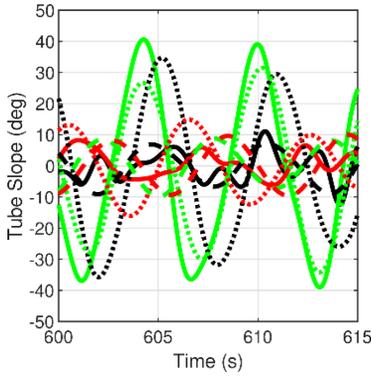
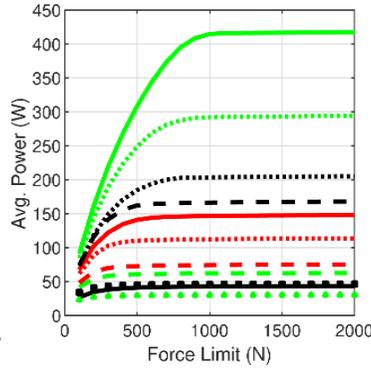
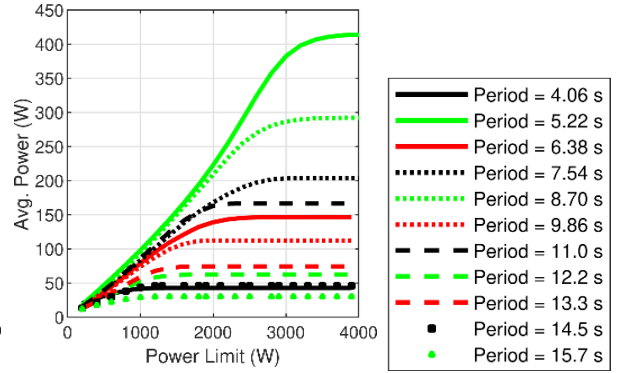

Figure 3: Tube slopes.  Figure 4: Parametric sweep of peak force limit.  Figure 5: Parametric sweep of peak power limit.

TABLE II AVERAGE POWER OUTPUT FOR DIFFERNET PTO STRATEGIES

|  | Average output power (W) | | |
|---|---|---|---|
| **Peak period (s)** | **Passive Damping** | **Reactive Damping** | **Discrete Damping** |
| **4.06** | 17.78 | 38.88 | 42.8 |
| **5.22** | 362.13 | 367.4 | 383.4 |
| **6.38** | 91.96 | 109.5 | 146.55 |
| **7.54** | 141.86 | 154.7 | 203.36 |
| **8.70** | 222.69 | 232.8 | 286.73 |
| **9.86** | 64.80 | 86.55 | 112.48 |
| **11.0** | 106.27 | 127 | 166.37 |
| **12.2** | 30.55 | 51.14 | 62.08 |
| **13.3** | 37.67 | 61.05 | 74.34 |
| **14.5** | 19.66 | 42.74 | 45.82 |
| **15.7** | 13.02 | 30.86 | 30.14 |

### B. Generator Ratings

Figs. 4 and 5 show the impact of the generator peak force and power limits on the average power that can be captured in each sea state using the discrete PTO strategy. Increasing the force and power limits increases the average power that can be captured for each of the different sea states. However, increasing these limits beyond 1000 N and 3000 W, respectively, yields diminishing returns, so these are selected as the force and power design targets for the linear generator.

### III. PMTL GENERATOR DESIGN

#### A. Generator Design Approach and Parameters

Fig. 2 shows the architecture of the generator. It consists of back iron, radially magnetized magnets, and outer windings on the stator. A similar design is proposed in [9] and [10] in which an analytical solution has been represented for the design of such generators. However, in this study, the design analysis is done using parametric finite element analysis (FEA) simulations of the generator in ANSYS Maxwell. Due to the simplicity of the design and its symmetry around the axis, parametric 2D simulations are used to rapidly characterize its performance. The design parameters and the acceptable range for each one is listed in Table III. Based on these ranges, all the cases are generated and simulated. There are two constraints for the design of the generator: 1) The minimum acceptable outer radius for the shaft is set to 50 mm so it can withstand the translator weight without significant deflection 2) The total outer radius of the generator ($R_e$) should be less than 105 mm to fit inside the SR-WEC. Therefore, the cases that do not satisfy these two constraints are not considered. A total of 8160 cases were simulated for this study.

#### B. Optimization of the Generator Design

Based on the PTO study, the generator requires a force rating of 1000 N and a power rating of 3000 W. Parametric magnetostatic simulations are used to characterize the impacts of the design parameters, and transient simulations are performed to check the back-emf and force ripple of the best designs. The airgap is assumed to be 1 mm, and a rms current density of 5 A/mm$^2$ at peak force is assumed for the windings.

TABLE III PMTL GNERATOR PARAMETERS

| Design Parameter | Range |
|---|---|
| **Shaft Radius** | 50 mm |
| **Magnet thickness ($T_m = R_m - R_0$)** | 2-10 mm |
| **Back iron thickness** | 5-25 mm |
| **Translator length ($l_e$)** | 100-300 mm |
| **Translator poles** | 2-12 |
| **Winding thickness ($R_s - R_i$)** | 10-30 mm |
| **Airgap (G)** | 1mm |
| **Stator Yoke** | 5 mm |
| **Copper Fill Factor** | 75% |

Fig. 6 show the effect of the winding thickness for different values of magnet thicknesses, which indicates the effects of magnetic loading and electric loading on the generator force. The results shown in Fig. 6 are the highest translator force for each magnet thickness and winding thickness while letting the translator length, pole count and back iron thickness vary freely over the range of values mentioned in Table III. As seen in Fig. 6, the force plateaued after a certain winding thickness due to the reduced flux density in the outer turns. Additionally, as the thickness of PM and winding is increased, the back iron thickness reduces to meet the 105 mm outer radius constraint, which reduces the force for the highest winding and magnet thicknesses. However, generally, the increase in magnet thickness increases the force for various values of coil thickness because increasing the magnet thickness increases the flux density, as well as increasing the air gap radius. In order to minimize the cost, a magnet thickness of 4 mm and a winding thickness of 5 mm are selected.

Fig. 7 shows the effect of the translator length for pole counts with magnet and winding thicknesses of 4 mm and 5 mm respectively. The minimum and maximum allowable lengths for the translator are 100mm and 300mm, respectively. The pole pitch, number of poles and translator length need to be determined. Referring to Fig. 7, a translator length of 300 mm with 8 poles satisfies the 1 kN force requirement.

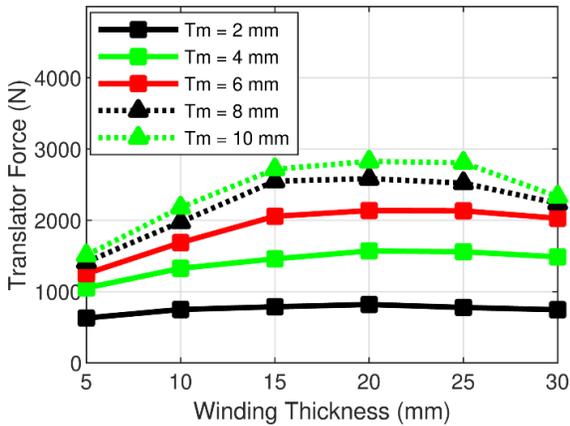

Figure 6: Force vs winding thickness for different magnet thicknesses

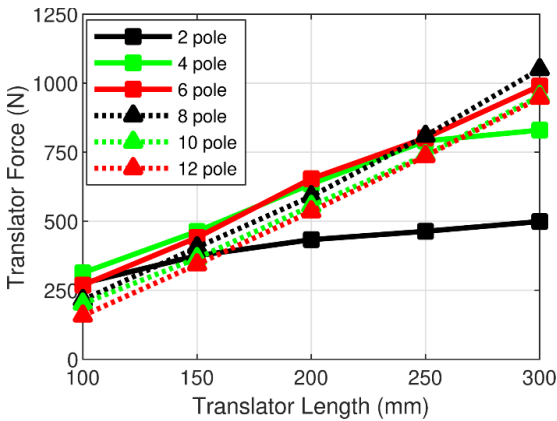

Figure 7: Force vs translator back iron thickness for different stator yoke thicknesses

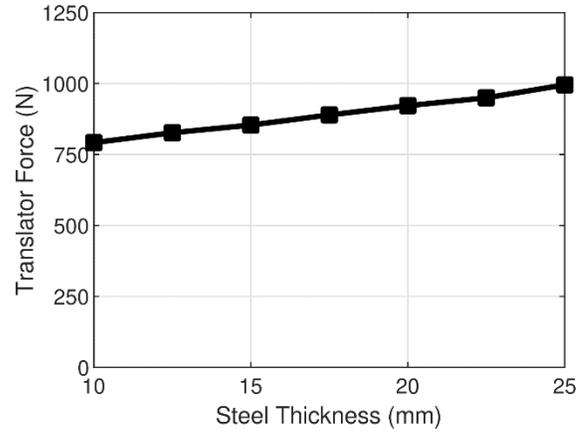

Figure 8: Force vs translator back iron thickness

Next, the thickness of the back iron is selected. According to Fig. 8, increasing the back iron thickness increases the force. With the given constrains, a back iron thickness of 25mm is chosen. The thickness of the stator yoke does not significantly affect the force production. However, removing the stator yoke completely does result in a force reduction of about 14%. Consequently, the back iron and stator thickness of 25 mm and 5 mm are selected, respectively.

*C. Winding Design*

The number of turns for each winding is determined to produce the Ampere-turns according to the designed winding area. As shown in Fig. 9, there is one coil per phase per pole. With a current density of 5 A/mm$^2$, winding thickness of 5 mm and pole pitch of 37.5 mm, each coil is designed to have 90 turns of 20 AWG wire. The force ripple predicted by transient simulation is less than 4% of the average force which is acceptable for this application. The final specifications of the generator are summarized in Table IV.

IV. EXPERIMENTAL PROTOTYPE

A smaller prototype with similar architecture is designed and fabricated to investigate the performance of the PMTL generator inside the SR-WEC. The prototype is cautiously designed and serves as a proof of concept prototype. Table V lists the parameters and dimensions of the prototype.

The translator is assembled using commercially available N42 magnets with a total of 12 poles. The back iron inner radius is 6 mm and its outer radius is 31.75 mm. The pole pitch is 19.05 mm. The magnet thickness is 6.35 mm and each pole consist of 4 arc magnets. Fig. 10 shows the translator.

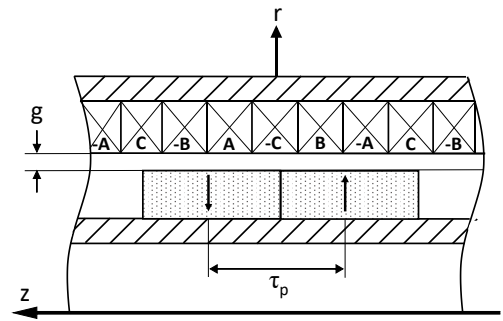

Figure 9: Winding Design from [10]

TABLE IV  PMTL GNERATOR SPECS

| Design Parameter | Value |
|---|---|
| $R_0$ | 75 mm |
| $R_m$ | 79 mm |
| $R_i$ | 80 mm |
| $R_s$ | 85 mm |
| $R_e$ | 90 mm |
| $l_e$ | 300 mm |
| Translator poles | 8 |
| G | 1mm |
| $\tau_p$ | 37.5 mm |
| Translator Back Iron | 25 mm |
| Stator Yoke | 5 mm |
| Turns per Coil | 90 |
| Magnet Type | N50 |

TABLE V  PMTL SMALL PROTOTYPE SPECS

| Design Parameter | Value |
|---|---|
| $R_0$ | 31.75mm |
| $R_m$ | 38.1mm |
| $R_i$ | 41.1mm |
| $R_e$ | 51.1mm |
| $l_e$ | 228.6mm |
| Translator Poles | 12 |
| G | 3mm |
| $\tau_p$ | 19.05mm |
| Translator Back Iron | 23.25mm |
| Turns per Coil | 70 |
| Magnet Type | N42 |

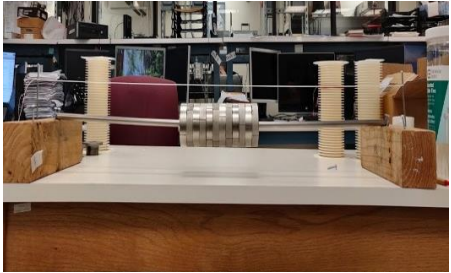

Figure 10: Translator of the prototype

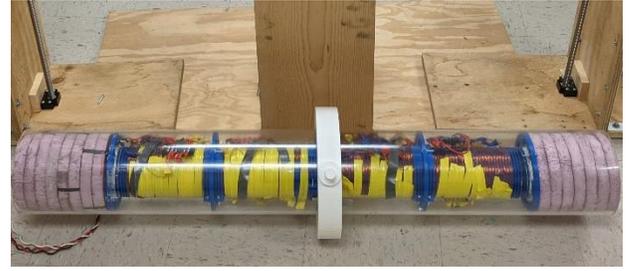

Figure 11: Assembled prototype

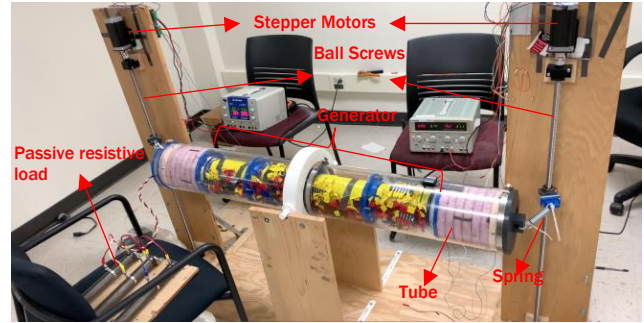

Figure 12: Testbed

The stator is fabricated using additive manufacturing. The total stator length is 914.4 mm, so the translator can travel a maximum distance of 685.8 mm inside the stator. Fig. 11 shows the assembled stator inside a tube that represents the SR-WEC. Each coil has of 70 turns of 20 AWG wires all connected in series. FEA simulations show that the prototype is capable of producing 140 N at its rated peak current of 3.66A.

A testbed was built to emulate the wave motion, as shown in Fig. 12. It includes two stepper motors, ball screws, and springs. By controlling the motors according to wave frequency and amplitude, the prototype can be tilted with different frequencies and slopes to evaluate the generator for different emulated sea states.

The 22 stator windings can be connected in series or parallel. Series connection results in high resistance and low efficiency. However, parallel connection can result in circulating currents. In the initial stages, the generator is tested with series connection of windings due to its simplicity. Measurements with LCR meter show an inductance of 44 mH and resistance of 67.2 Ω measured line-line with the series connection.

In the initial stages, only passive PTO damping is used. The generator terminals are connected to a three-phase resistive load with 33.6 Ω in each phase. Therefore, 50% of the generated power by the generator is dissipated in the windings due to series connection of the windings.

Fig. 13, shows phase voltage, current, and instantaneous three-phase output power for a sliding angle of 40°. The phase current and terminal voltage peak at 18.8 V and 0.57 A, respectively. At 40°, the total three-phase average power goes up to 16 W immediately before the translator reaches the end of the stator around 1.1 s.

Fig. 14, shows the generator phase voltage, current, and instantaneous three-phase output power for a sliding angle of 50°. At this angle, the translator speed is higher when it reaches the end of the stator around 0.95 s, so the peak current, voltage and power are higher as well. The peak voltage and current are 23.69 V and 0.75 A, respectively. The total three-phase average power goes up to about 28 W.

As stated previously, 50% of the generator power is dissipated in the stator windings due to the large resistance of the series connection and the selection of the load resistance. Moreover, passive PTO does not extract the maximum power from the SR-WEC. An active or discrete PTO strategy or parallel connection of stator windings could result in higher output power.

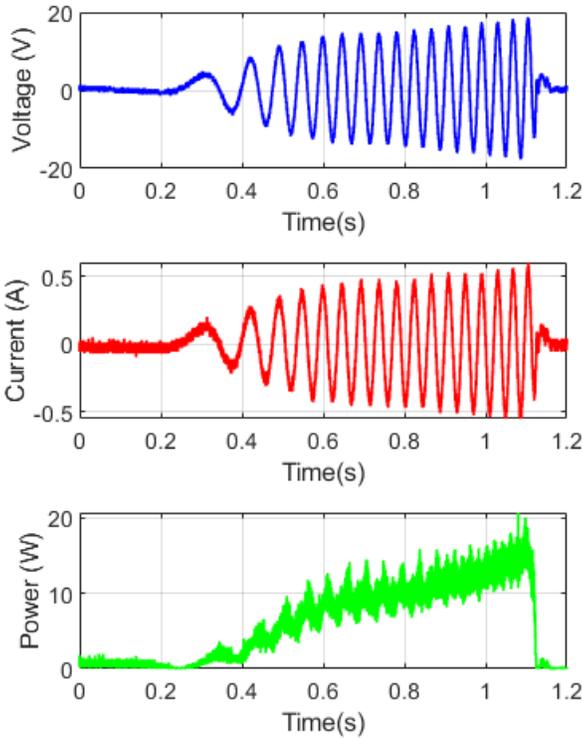

Figure 13: Generator line-to-neutral voltage, current, and three-phase power outputs at 40° sliding angle

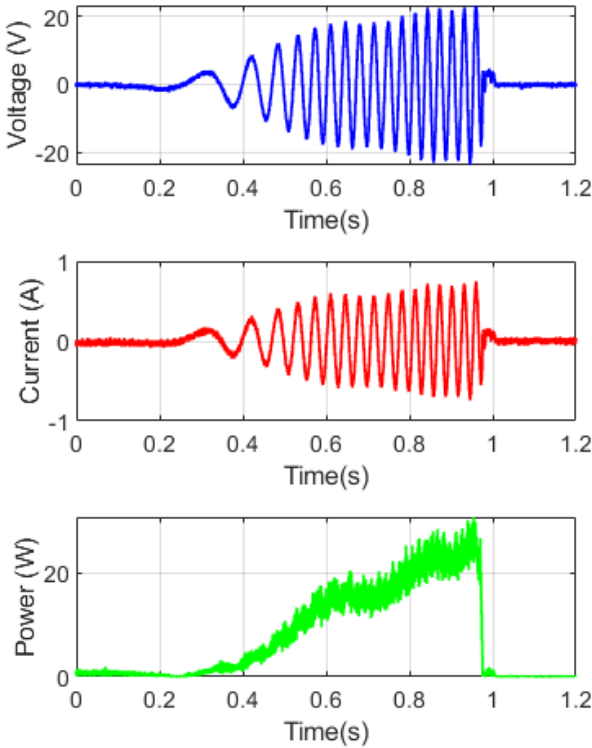

Figure 14: Generator line-to-neutral voltage, current, and three-phase power outputs at 50° sliding angle

## V. CONCLUSION AND FUTURE WORK

In this work the design of a 1000 N, 3 kW PMTL generator for SR-WEC was discussed. The generator is slotless to eliminate cogging force. Parametric FEA simulation was performed to find the optimized design. An experimental prototype with similar architecture is designed and fabricated to evaluate the performance of the generator. Initial tests and results of the experimental prototype were presented.

As future work, the authors intend to implement the reactive and discrete PTO strategies mentioned in section II. In addition, the generator will be controlled using a 3-phase PWM rectifier using sensorless control methods [16]-[19].

## ACKNOWLEDGMENT

This research was financially supported by U.S. Department of Energy Water Power Technologies Office (DE-EE0008630). Part of this research was conducted using advanced computing resources provided by Texas A&M High Performance Research Computing. The authors would like to thank ANSYS for their support of EMPE Lab by providing FEA software. The authors would like to thank Department of Energy for supporting this work. The authors would also like to thank Dr. Matthew Johnson of the U.S. Army Research Laboratory for his help in setting up the FEA simulations.

## REFERENCES

[1] US. Department of Energy, "Quadrennial technology review 2015," Washington, DC, USA 2015.
[2] H. Kang and M. Kim, "Method and apparatus for wave energy conversion," US Patent 10,352,290, 2019.
[3] C. Jin, H. Kang, M. Kim, and F. P. Bakti, "Performance evaluation of surface riding wave energy converter with linear electric generator," *Ocean Eng.*, vol. 218, pp. 108141, Dec. 2020.
[4] N. Hodgins, O. Keysan, A. S. McDonald, and M. A. Mueller, "Design and Testing of a Linear Generator for Wave-Energy Applications," *IEEE Trans. Ind. Electron.*, vol. 59, no. 5, pp. 2094-2103, May 2012.
[5] A. Musolino, M. Raugi, R. Rizzo, and L. Sani, "A Semi-Anaytical Model for the Analysis of a Permanent Magnet Tubular Linear Generator," *IEEE Trans. Ind. Appl.*, vol. 54, no. 1, pp. 204-212, Jan.-Feb. 2018.
[6] L. Huang, M. Hu, Z. Chen, H. Yu, and C. Liu, "Research on a Direct-Drive Wave Energy Converter Using an Outer-PM Linear Tubular Generator," *IEEE Trans. Magn.*, vol. 53, no. 6, pp. 1-4, Jun. 2017.
[7] L. Cappelli et al., "Linear Tubular Permanent-Magnet Generators for the Inertial Sea Wave Energy Converter," *IEEE Tran. Ind. Appl.*, vol. 50, no. 3, pp. 1817-1828, May-Jun. 2014.
[8] J. Prudell, M. Stoddard, E. Amon, T. K. A. Brekken, and A. von Jouanne, "A Permanent-Magnet Tubular Linear Generator for Ocean Wave Energy Conversion," *IEEE Trans. Ind. Appl.*, vol. 46, no. 6, pp. 2392-2400, Nov.-Dec. 2010.
[9] J. Wang, G. W. Jewell, and D. Howe, "Design optimisation and comparison of permanent magnet machines topologies," *IEE. Proc. Elect. Power Appl.*, vol. 148, pp. 456–464, Sep. 2001.
[10] J. Wang, G. W. Jewell, and D. Howe, "A general framework for the analysis and design of tubular linear permanent magnet machines," *IEEE Trans. Magn.*, vol. 35, no. 3, pp. 1986-2000, May 1999.
[11] C. Liu, H. Yu, M. Hu, Q. Liu, and S. Zhou, "Detent Force Reduction in Permanent Magnet Tubular Linear Generator for Direct-Driver Wave Energy Conversion," *IEEE Trans. Magn.*, vol. 49, no. 5, pp. 1913-1916, May 2013.
[12] J. Zhang, H. Yu, M. Hu, L. Huang and T. Xia, "Research on a PM Slotless Linear Generator Based on Magnet Field Analysis Model for Wave Energy Conversion," *IEEE Trans. Magn.*, vol. 53, no. 11, pp. 1-4, Nov. 2017.

[13] E. Tedeschi and E. Molinas, "Control Strategy of Wave Energy Converters Optimized under Power Electronics Rating Constraints," in *Proc. Int. Conf. Ocean Energy*, 2010, pp. 1-6.

[14] R. So, M. Starrett, K. Ruehl, and T. K. A. Brekken, "Development of control-Sim: Control strategies for power take-off integrated wave energy converter," in *Proc. IEEE Power & Energy Soc. Gen. Meet.*, 2017, pp. 1-5.

[15] National Oceanic and Atmospheric Administration. [Online]. Available: https://www.ndbc.noaa.gov/station_history.php?station=41002

[16] S. Bolognani, R. Oboe, and M. Zigliotto, "Sensorless full-digital PMSM drive with EKF estimation of speed and rotor position," *IEEE Trans. Ind. Electron*, vol. 46, no. 1, pp. 184-191, Feb. 1999.

[17] A. M. Alshawish, S. M. Eshtaiwi, A. Shojaeighadikolaei, A. Ghasemi, and R. Ahmadi, "Sensorless Control for Permanent Magnet Synchronous Motor (PMSM) Using a Reduced Order Observer," *2020 IEEE Kansas Power and Energy Conf.*, 2020, pp. 1-5

[18] W. Zhao, S. Jiao, Q. Chen, D. Xu, and J. Ji, "Sensorless Control of a Linear Permanent-Magnet Motor Based on an Improved Disturbance Observer," *IEEE Trans. Ind. Electron.*, vol. 65, no. 12, pp. 9291-9300, Dec. 2018.

[19] H. A. Hussain and H. A. Toliyat, "Back-EMF based sensorless vector control of tubular PM linear motors," in *Proc. IEEE Int. Elect. Mach. Drives Conf.*, 2015, pp. 878-883.